**National-scale electricity peak load forecasting:**

**Traditional, machine learning, or hybrid model?**

Juyong Lee and Youngsang Cho*

Juyong Lee, Ph.D.

Postdoctoral Research Fellow

Department of Industrial Engineering, College of Engineering, Yonsei University, 50 Yonsei-Ro, Seodaemun-gu, Seoul, 03722, South Korea

Tel: +82-2-2123-7913

E-mail: dlwndyd@yonsei.ac.kr

Youngsang Cho, Ph.D. (*Corresponding author)

Professor

Department of Industrial Engineering, College of Engineering, Yonsei University, 50 Yonsei-Ro, Seodaemun-gu, Seoul, 03722, South Korea

Tel: +82-2-2123-5727

Fax: +82-2-364-7807

E-mail: y.cho@yonsei.ac.kr

# National-scale electricity peak load forecasting:

# Traditional, machine learning, or hybrid model?

## Abstract

As the volatility of electricity demand increases owing to climate change and electrification, the importance of accurate peak load forecasting is increasing. Traditional peak load forecasting has been conducted through time series-based models; however, recently, new models based on machine or deep learning are being introduced. This study performs a comparative analysis to determine the most accurate peak load-forecasting model for Korea, by comparing the performance of time series, machine learning, and hybrid models. Seasonal autoregressive integrated moving average with exogenous variables (SARIMAX) is used for the time series model. Artificial neural network (ANN), support vector regression (SVR), and long short-term memory (LSTM) are used for the machine learning models. SARIMAX-ANN, SARIMAX-SVR, and SARIMAX-LSTM are used for the hybrid models. The results indicate that the hybrid models exhibit significant improvement over the SARIMAX model. The LSTM-based models outperformed the others; the single and hybrid LSTM models did not exhibit a significant performance difference. In the case of Korea's highest peak load in 2019, the predictive power of the LSTM model proved to be greater than that of the SARIMAX-LSTM model. The LSTM, SARIMAX-SVR, and SARIMAX-LSTM models outperformed the current time series-based forecasting model used in Korea. Thus, Korea's peak load-forecasting performance can be improved by including machine learning or hybrid models.

# 1. Introduction

Electricity peak load forecasting is essential for stable and efficient power system operation, power generation plans, and supply and demand management. Recently, the electricity load demand has increased, owing primarily to economic growth, electrification, and climate change [1–3]. A failure to accurately forecast peak load can negatively impact the national economy and finances owing to the additional costs of unexpected and surplus power generation [4,5]. Therefore, one of the main goals of power suppliers and system operators is to accurately forecast consumers' electricity demand. Information obtained from demand forecasting is important for (1) short-term generator preventive maintenance schedules, (2) daily/weekly/annual generator shutdown schedules, and (3) economical and stable system operation.

In the power industry, peak load forecasting has primarily focused on buildings or specific regions [5–8]. Forecasting the peak loads for small areas enables the selection of different exogenous factors, which can be flexibly and quickly reflected in the future strategies and policies based on the analysis results [9]. However, an accurate national-scale forecasting is essential for long-term energy planning to address global climate change and resource issues. Numerous forecasting models have been studied globally with the improvement of computer processing power and advances in artificial intelligence (AI), and different models, including single, ensemble, and hybrid, have been proposed and utilized. Many countries have their own forecasting models and prepare for future uncertainties through scenario analysis based on their forecasts.[1] In South Korea, power generation and supply are managed by a government-affiliated organization, Korea Power Exchange (KPX), which uses a short-term load forecaster (KSLF) to predict the electricity demand for the next day and uses this prediction to operate the power system.

---

[1] Institutes and companies such as Entergy, New York Independent System Operator, and the City of College Station in the United States, Gestionnaire du Reseau de Transport d'Electricité in France, the Independent Electricity System Operator in Canada, and the Institute of Electrical Engineers of Japan in Japan, are developing and utilizing demand forecasting models [10,11].

Models that forecast the electricity peak load are primarily divided into time series models, based on econometrics, and machine (or deep)-learning models, based on AI algorithms [12,13]. When the electricity market conditions move with a constant trend, both the time series and machine learning models exhibit acceptable predictive power [14]. However, if the market changes rapidly and nonlinearly, the time series models cannot provide an accurate forecast. This is because general time series models are formulated as a linear combination of their own values from the previous time-step. Machine learning models that can process nonlinear data offer advantages in predictive power over time series models [15–17]. In this regard, the machine learning models can be used as auxiliary models or next-generation models by complementing or replacing existing time series models. Moreover, a hybrid model, using both time series and machine learning models, was proposed by Zhang [18], and different follow-up studies have been actively applied to peak load forecasting.

Table 1 presents a brief description of the previous studies on electricity peak load forecasting. There have been studies comparing the predictive powers of the different time series models over the past decade [19–21]. They have indicated that without exogenous variables, the Holt–Winters model, based on exponential smoothing, outperforms the autoregressive integrated moving average (ARIMA), seasonal ARIMA (SARIMA), and generalized autoregressive conditional heteroskedasticity (GARCH) models. However, considering exogenous variables, such as temperature, humidity, rainfall, weekends, and holidays, significantly improves the performance of forecasting models; thus, ARIMA with exogenous variables (ARIMAX) or SARIMA with exogenous variables (SARIMAX) have been more actively utilized than the Holt–Winters model or other time series models. Based on previous studies on time series models, the SARIMAX model is the most appropriate for peak load forecasting; the Holt–Winters model does not typically process exogenous variables (or external regressors).

**Table 1.** Summary of related previous studies.

| Author(s) | Year | Main Variables | Model(s) | Description |
|---|---|---|---|---|
| Park et al. [22] | 1991 | peak load | artificial neural network (ANN) | · The first study to apply artificial neural networks to electricity demand forecasting.<br>· ANN has a lower error than traditional time series predictions. |
| Desouky and Kateb [23] | 2000 | peak load, temperature | ARIMA, ANN, hybrid | · ANN captures the nonlinear relationship between the peak load and temperature.<br>· Both ANN and hybrid models outperform the ARIMA model. |
| Taylor [20] | 2010 | peak load, holiday | ARIMA, Holt–Winters | · A three-season Holt–Winters model has superior predictive power compared to the ARIMA model. |
| Lee et al. [19] | 2013 | peak load, seasons | ARIMA, Holt–Winters, GARCH | · The Holt–Winters model demonstrates the best predictive power owing to Korea's distinct seasonality. |
| Ji et al. [24] | 2013 | peak load, temperature | extreme learning machine (ELM) | · ELM does not execute repetitive learning, unlike other neural network models; thus, the learning time is significantly faster than typical repetitive learning models.<br>· Peak load is more correlated with the load data of the previous seven days than with temperature variables. |
| Jung and Kim [21] | 2014 | peak load, seasonal components, temperature | ARIMA, GARCH | · The GARCH model has demonstrated greater predictive power than the ARIMA model owing to high peak load volatility.<br>  Considering cooling degree day (CDD) and heating degree day (HDD), rather than only temperature, improves the predictive power of the models. |

| | | | | | |
|---|---|---|---|---|---|
| Tak et al. [25] | 2016 | peak load, weather | support vector regression (SVR) | · | Predictive power is improved by more than 15% when using pseudo-temperature variables. |
| Chen et al. [26] | 2017 | peak load, temperature | SVR | · | Predictive power is improved with a lag of two hours in temperature variables. |
| Vu et al. [27] | 2017 | peak load, holiday, DST | Autoregressive (AR), ARIMA, autoregressive time varying (ARTV), periodic autoregressive moving average, ANN | · | The ARTV model outperforms other time series models; thus, including a time-varying component in an autoregressive model can improve its predictive power. |
| Lee and Kim [28] | 2019 | peak load, temperature, holiday | ARIMA, ANN | ·<br>· | The neural network model has a greater predictive power than the ARIMA model.<br>Exogenous variables improve the predictive power of ARIMA models. |
| Cai et al. [29] | 2019 | peak load, temperature | ARIMA, RNN, convolutional neural network (CNN) | · | The predictive power of the gated CNN model outperforms that of the seasonal ARIMAX model by 22.6%. |
| Zhang et al. [30] | 2017 | peak load | ARIMA, spectrum analysis, support vector machine (SVM), ANN, hybrid | · | The hybrid model (spectrum analysis and SVM) outperforms other time series and SVM models. |
| Chou and Tran [31] | 2018 | peak load, temperature, weekends | ARIMA, ANN, SVM, hybrid | ·<br>· | ANN has the greatest predictive power among single models.<br>The seasonal ARIMA-SVM hybrid model outperforms all the other models. |
| Muzaffar and | 2019 | peak load, | ARIMA, long short-term | · | The LSTM model outperforms all other time series models |
| Afshari [32] | | temperature | memory (LSTM) | | (ARIMA, SARIMA, SARIMAX). |
| Guo et al. [33] | 2018 | peak load, weather | multilayer perceptron | · | Adding meteorological features (temperature, wind speed, humidity) improves the predictive power of the LSTM model. |
| Sheng and Jia [34] | 2020 | peak load | SARIMAX, LSTM, SARIMAX-LSTM | · | SARIMAX–LSTM outperforms all other single models. |

Both time series models and machine learning methods have been actively utilized for peak load forecasting. Furthermore, machine learning models have been used to forecast electricity price and power generation, which are interrelated with the electricity demand and peak load [35–37]. Since Park et al. [22] proposed an ANN-based peak load-forecasting model in 1991, numerous researchers have applied and proposed machine learning models and traditional time series models. The most widely used model is the ANN-based model; support vector machine (SVM)-based models have also been actively utilized in peak load forecasting. Lee and Kim [28] and Desouky and Kateb [23] compared the ANN model with the traditional ARIMA model; the ANN model demonstrated a significantly greater predictive power than the ARIMA model because the ANN model could capture the nonlinear characteristics of the peak load data. Vu et al. [27] compared time series models with the ANN model and found that an autoregressive-based time varying model outperformed the ANN model. Cai et al. [29], Muzaffar and Afshari [32], and Guo et al. [33] utilized deep neural network models, such as recurrent neural network (RNN), convolutional neural network (CNN), and long short-term memory (LSTM), and found that these models outperformed traditional time series models. In addition to the neural network models, support vector models such as SVM and support vector regression (SVR) have been applied to peak load forecasting in several studies. Chen et al. [26] and Tak et al. [25] applied SVR with meteorological variables, demonstrating significantly improved predictive power.

Hybrid models, which combine elements of time series and machine learning models, have recently been applied to peak load forecasting. In the ARIMA–ANN hybrid model, ANN captures the nonlinear components of the peak load data, which the ARIMA-based models have failed to do. Zhang et al. [30] and Nie et al. [38] compared ARIMA, SVM, and hybrid models, where the ARIMA–SVM hybrid model had the greatest predictive power. Chou and Tran [31] demonstrated that ANN outperforms SVM, as well as different single models including time series and machine learning. However, SVM improves the predictive power of the hybrid models more than ANN; thus, ARIMA–SVM outperformed all the models studied. Sheng and Jia [34] forecasted the peak load of a city in China using SARIMAX, LSTM, and SARIMAX–LSTM and demonstrated that SARIMAX–LSTM had the greatest predictive power. The majority of previous studies concluded that hybrid models outperform other single time series- or machine learning-based models.

To summarize the previous studies: (1) weekends, holidays, and meteorological variables such as temperature, humidity, rainfall, and wind speed are significantly correlated with the peak load; these should be included when modeling the accurate forecasting of electricity peaks; (2) machine learning models typically outperform time series models; however, the best machine learning model has varied from study to study; and (3) hybrid models exhibit greater predictive power than single models.

There have been several studies comparing forecasting models and identifying the top-performing models for electricity peak load forecasting. However, the majority of these have targeted buildings or specific regions. No study, applying traditional or recently developed hybrid models, has comprehensively focused on the peak load forecasting of an entire country. Thus, this study can offer more comprehensive and long-term directions in terms of methodology and policy than previous studies through the comparative analysis of different country-level forecasting models.

The main purpose of this study is twofold: (1) to find the most accurate predictive model among time series, machine learning, and hybrid approaches as well as to verify if the predictive power of hybrid models exhibits a significant improvement over existing single models and (2) to verify that the model used in this study is superior to the model currently used by KPX as well as to derive the implications for electricity supply and demand regarding reserve margins. Furthermore, directions for the improvement of Korea's current peak load-forecasting model are also suggested. The remainder of this study is organized as follows: Section 2 present the models and data utilized. The forecasting results of the models and several implications are described in Section 3. Conclusion and limitations of the study are presented in Section 4.

## 2. Methodology

This study forecasts the electricity peak load by applying different models, including traditional time series, machine learning, and hybrid models. Although different detailed and advanced models from both time series and machine learning have been actively proposed, this study compares the predictive powers of the models that have primarily been utilized in previous studies. Thus, this study uses the SARIMAX model for a time series model, and ANN, SVR, and LSTM as machine learning models. The following subsections briefly describe the models used in this study and further consider the hybrid models that combine the time series and machine learning models.[2] Moreover, this study describes measurement statistics to evaluate the models' predictive power. Figure 1 presents a flowchart of the research steps.

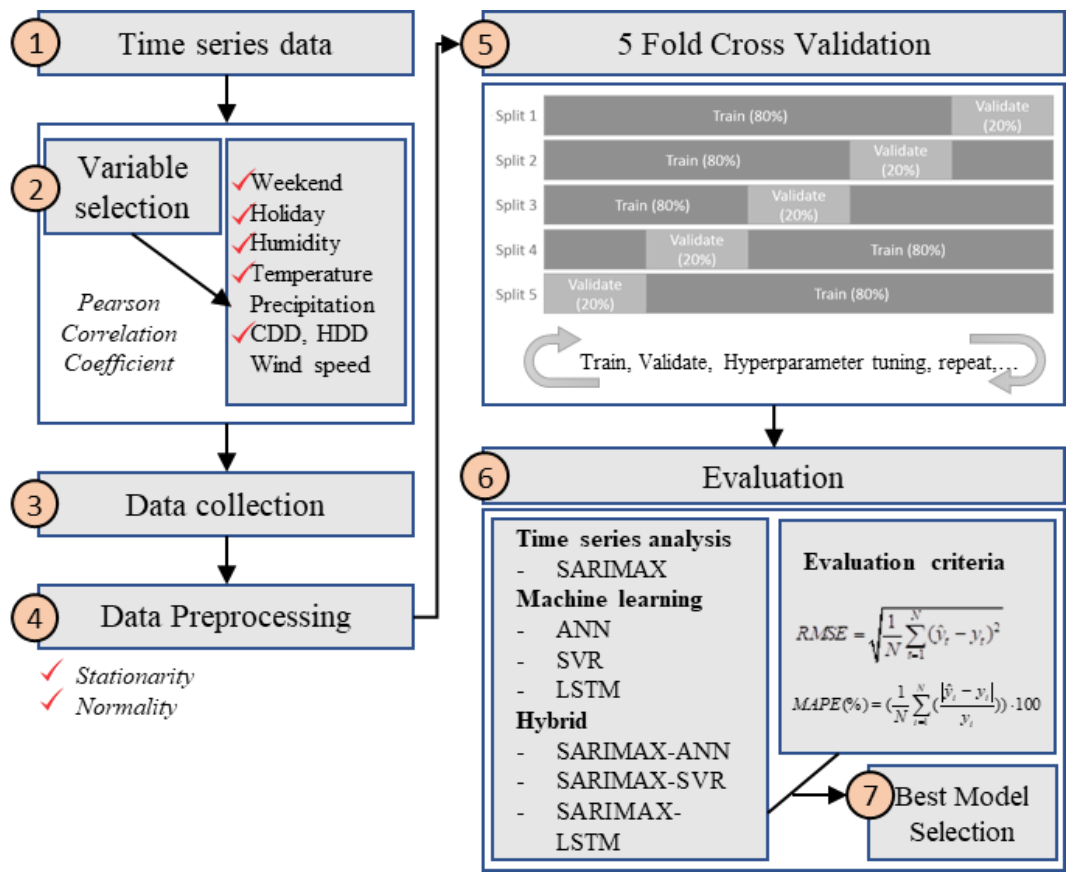

**Fig. 1.** Flowchart of research steps.

---

[2] The detailed algorithms or descriptions of the models can be found in the previous studies cited in this section.

## 2.1. SARIMAX

This study utilizes the SARIMAX model, which is one of the most widely used models for time series data. The ARIMA model, also called the Box–Jenkins model, named after the researchers who introduced it, is a method of analyzing nonstationary time series data [39]. The SARIMA model is applied to a nonstationary time series, whereas seasonality exists in the ARIMA model. The SARIMAX model considers exogenous variables in addition to the SARIMA model. The equation for SARIMAX $(p, d, q)(P, D, Q)_S$ generating time series $Y_t$ with the mean $\mu$, can be expressed as

$$\phi_p(B)\Phi_P(B^S)(1-B^S)^D(Y_t - \mu) = \theta_q(B)\Theta_Q(B^S)\varepsilon_t + \sum_{i=1}^{k}\gamma_i x_{it}, \tag{1}$$

where $Y_t$ and $\varepsilon_t$ are the observations and random errors at time $t$, respectively, and $\gamma_i$ is the coefficient of the exogenous variable $x_{it}$; $p$ is the order of the AR process, $q$ is the order of the moving average (MA) process, $P$ is the order of the seasonal AR process, and $Q$ is the order of the seasonal MA process; $B$ is the back-shift operator, $d$ is the difference order, and $S$ is the seasonal difference order. In addition, $\phi_p(B)$, $\theta_q(B)$, $\Phi_P(B^S)$, and $\Theta_Q(B^S)$ represent polynomials as follows:

$$\begin{aligned}
\phi_p(B) &= 1 - \phi_1 B - ... - \phi_p B^p, \\
\theta_q(B) &= 1 - \theta_1 B - ... - \theta_q B^q, \\
\Phi_P(B^S) &= 1 - \Phi_1 B^S - ... - \Phi_P B^{PS}, \\
\Theta_Q(B^S) &= 1 - \Theta_1 B^S - ... - \Theta_Q B^{QS}.
\end{aligned} \tag{2}$$

$p$, $d$, $q$, $P$, $D$, and $Q$ are typically determined using two techniques: (1) plotting a correlogram and observing an autocorrelation function (ACF) and partial ACF (PACF) and (2) using all possible combinations of $p$, $d$, $q$, $P$, $D$, and $Q$, and then selecting a combination to minimize the Akaike information criterion (AIC) or Bayesian information criterion (BIC) [40]. To reduce the time requirement of these processes, the *auto.arima* package in R has been widely used [41].

*2.2. ANN*

ANN is an algorithm designed to process data in a similar manner the human brain analyzes information. Neurons, the basic structural tissues of the brain, are interconnected to form a network called a neural network. An artificial network composed of structural units of a perceptron that mimics these neurons is called an ANN. A multilayer perceptron learns by adding hidden layers and a backpropagation algorithm between the input and output layers to enable nonlinear classification. The multilayer perceptron is typically more accurate for time series forecasting than a single-layer perceptron; therefore, it is one of the most widely utilized neural networks. The formula to determine the output value of each neuron in the hidden layer is as follows:

$$y_j(t) = \varphi(\sum_{i=1}^{N} w_{ij}(t)x_i(t) + w_{i0}) \,, \tag{3}$$

where $\varphi$ is an activation function that models a sigmoid function for the hidden layers, and inputs $x_i$ and outputs $y_i$ are interconnected with synaptic weights represented by $w_{ij}$ [42]. The weights are adjusted using a recursive algorithm with the outputs and error term $e_j$ as

$$w_{ij}(t+1) = w_{ij}(t) + e_j y_j(t) \,. \tag{4}$$

To process nonlinear time series data with exogenous variables when applying ANN-based models, a nonlinear autoregressive exogenous input (NARX) architecture is widely utilized [43–45]. Because the electricity peak load data is time series data and influenced by different external variables, the NARX architecture is appropriate for ANN fitting. Figure 2 displays the basic NARX architecture. In this architecture, the internal structure that performs the approximation of the mapping function is the multilayer perceptron.[3] ANN has excellent parallelism because each neuron acts as an independent processor. Because the information is distributed on numerous connection lines, it has a fault tolerance that does not influence the entire system, even if neurons experience problems [47,48].

---

[3] Boussaada et al. [46] explained the detailed algorithm, diagrams, and characteristics of the NARX architecture; please refer to their study for in-depth descriptions.

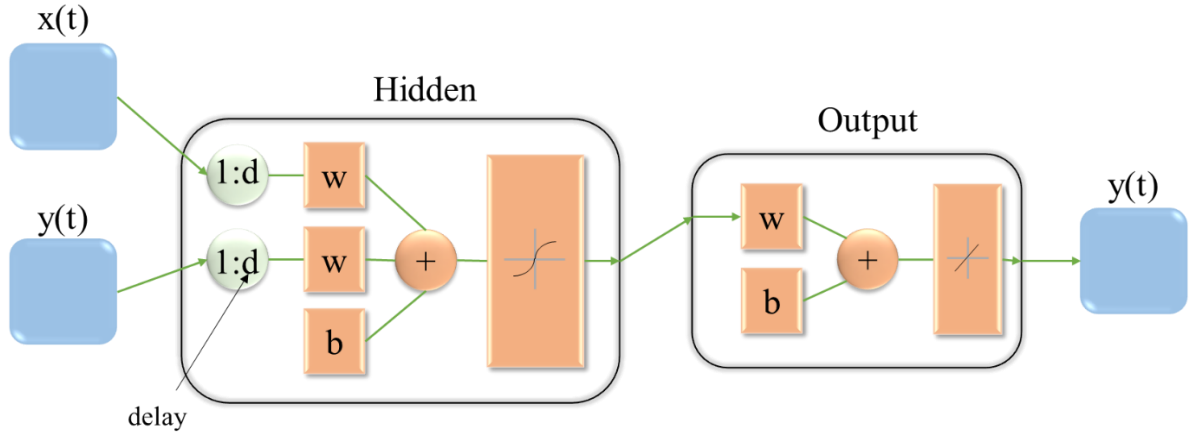

**Fig. 2.** NARX architecture.

## 2.3. SVR

SVM is a learning algorithm that conducting a binary classification in the field of pattern recognition; SVR is an extension model of SVM used for regression application [49,50]. The basic concept of SVR is to find the flattest regression function ensuring that the target variable of all training data is within a given deviation. The basic SVR schema can be expressed as shown in Figure 3. The regression function $f(x_t)$ is presented in equation (5); the optimization equation to minimize the error and find the flattest function can be expressed by the objective function in equation (6) with the constraints in equation (7):

$$f(x_t) = x_t{'}\beta + b \,,$$
(5)

$$minimize \quad S(\beta) = \frac{1}{2}\beta{'}\beta + C\sum_{t=1}^{T}(\xi_t + \xi^*_t) \,,$$
(6)

$$\textit{subject to} \quad \forall n : y_t - (x_t'\beta + b) \le \varepsilon + \xi_t$$
$$\forall n : (x_t'\beta + b) - y_t \le \varepsilon + \xi_t^* , \tag{7}$$
$$\forall n : \xi_t, \xi^* \ge 0,$$

where $C$ is the box constraint, which is a positive value that facilitates controlling the "penalty" imposed on observations outside the epsilon margin ($\varepsilon + \xi$) and prevent overfitting, $\frac{1}{2}\beta'\beta$ is the "minimum norm" that indicates the flatness of the function, and $T$ is the number of time series observations. As an advantage, SVR can be applied to not only linear regression functions but also nonlinear regression functions. Only the dot product of two vectors is required to solve the Lagrangian methods to obtain $f(x)$; thus, this allows the determination of the dot product of two vectors in high dimensions that can be represented linearly by kernel functions [51,52]. There are several kernel functions such as the radial basis, polynomial, and sigmoid functions [53]. Therefore, researchers should consider three hyperparameters for SVR model optimization: (1) box constraint, (2) epsilon margin, and (3) kernel functions. This study also conducted the optimization process with the three hyperparameters using a grid search.[4]

---

[4] A grid search is the process of finding the minimum error (best accuracy) hyperparameters by evaluating all possible combinations of the hyperparameter values based on cross-validation. This study conducted a grid search, as well as a randomized search and Bayesian optimization; however, the performance of the grid search was significantly better than that of the other methods.

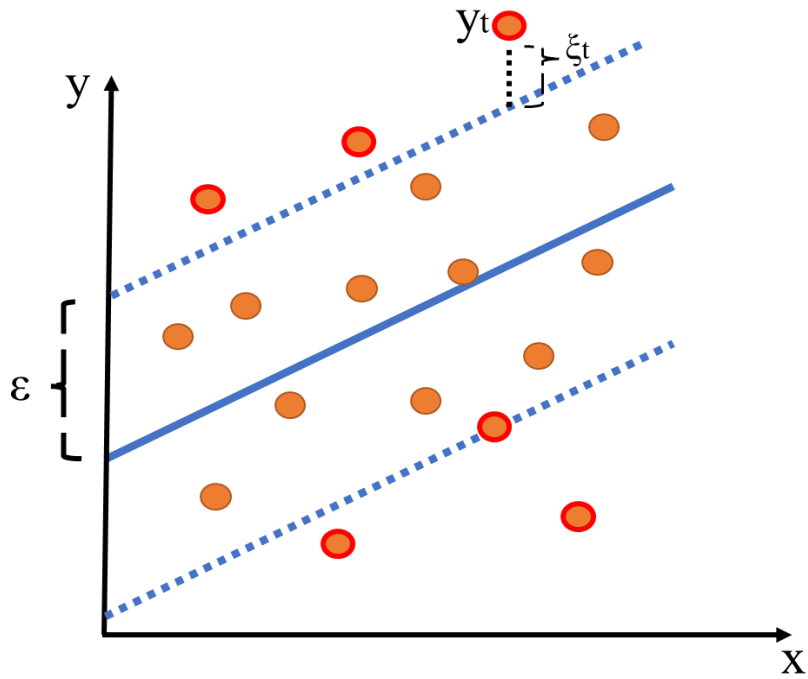

**Fig. 3.** Structure of SVR.

## *2.4. LSTM*

LSTM[5] is an RNN model proposed by Hochreiter and Schmidhuber [54], and it has been actively applied to the learning and forecasting of time series data. The traditional RNN is known to have a significantly lower gradient when backpropagating if the distance between the relevant information and point of using the information is large. This results in a significant decrease in learning performance, which is called the vanishing gradient problem [55,56]. LSTM is proposed to overcome this problem by adding a cell-state to the hidden-state of a typical RNN. Figure 4 presents the basic structure of the LSTM model. The memory cell state $c_{t-1}$, hidden state $h_{t-1}$, and input state $x_t$ interact with the LSTM gates such as input gate $i_t$, input node $g_t$, forget gate $f_t$, and output gate $o_t$. The formulations of the LSTM structure are presented as follows:

$$
\begin{aligned}
f_t &= \sigma(W_{fx}x_t + W_{fh}h_{t-1} + b_f) \\
i_t &= \sigma(W_{ix}x_t + W_{ih}h_{t-1} + b_i) \\
g_t &= \tanh(W_{gx}x_t + W_{gh}h_{t-1} + b_g) \\
o_t &= \sigma(W_{ox}x_t + W_{oh}h_{t-1} + b_o) \\
c_t &= f_t \odot c_{t-1} + i_t \odot g_t \\
h_t &= o_t \odot \tanh(c_t),
\end{aligned}
\tag{8}
$$

where $\odot$ is a Hadamard product (also called the elementwise multiplication); $\sigma$ represents the sigmoid activation function; and $W$ and $b$ are the weight matrices and bias vector parameters, respectively, which are learned during training. Thus, the network learns by updating, maintaining, or forgetting (erasing) the $t-1$ previous state information using the input, forget, and output gates. The LSTM model is one of the best-performing neural network-based models [57,58]. Recently, derivative models (the gated recurrent unit proposed by Cho et al. [59]) have been actively attempted and introduced. This study uses the basic LSTM model, which has been the most widely used model in previous studies.

---

[5] Lipton et al. [30] and Kong et al. [31] introduced the detailed structure and algorithm of the LSTM model. In this subsection, this study briefly introduces the basic concept and formulations of this model.

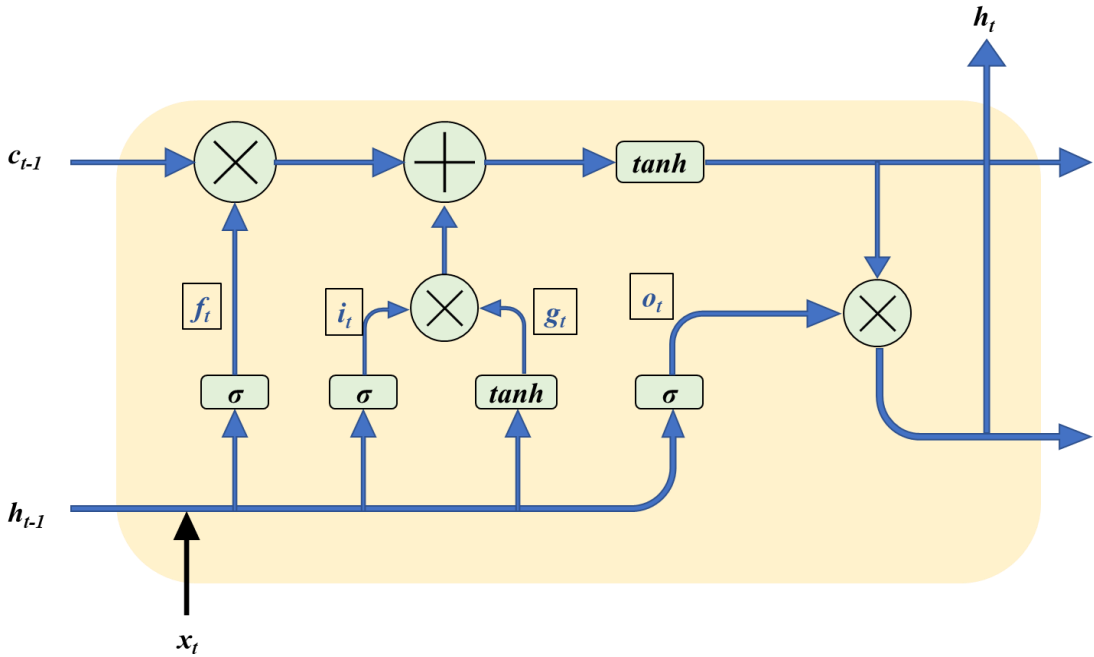

**Fig. 4.** Structure of LSTM.

## 2.5. Hybrid model

A hybrid model that combines a time series and machine learning model was proposed by Zhang [18]; subsequent studies have been performed to apply hybrid models in various fields including peak load forecasting. The hybrid model can be briefly expressed as

$$y_t = L_t + N_t, \tag{9}$$

where $L_t$ and $N_t$ are the linear and nonlinear functions, respectively, and $y_t$ represents the original value. The residual of time $t$, $r_t$ can be obtained by substituting the estimated values of the linear modeling process from the original values

$$r_t = y_t - \hat{L}_t, \tag{10}$$

where the linear modeling process can be conducted with time series models, and the residual can be estimated by machine learning models. Therefore, the estimated value of time $t$, $\hat{y}_t$ can be expressed as the sum of the estimated values of the time series and machine learning models, as follows:

$$y_t = \hat{L}_t + \hat{N}_t. \tag{11}$$

In fact, the term "hybrid model" in forecasting studies is not strictly limited to a combination of the time series and machine learning models. In studies that combine multiple types of machine learning or deep learning models to improve predictive power, the models are also referred to as hybrid models [60–62]. However, to prevent a misunderstanding, this study defines a "hybrid model" as a model based only on the concepts that have been described in this subsection.

## 2.6. Evaluation of the predictive power

Because this study compares the predictive powers of different forecasting models, a numerical criterion for performance comparison is required. The accuracy of the out-of-sample data is most commonly presented as the root mean squared error (RMSE) or the mean absolute percentage error (MAPE). Because RMSE is a scale-dependent error and MAPE is a percentage-dependent error, this study presents both the RMSE and MAPE of each model and compares their predictive power. The formulations of RMSE and MAPE are as follows:

$$RMSE = \sqrt{\frac{1}{N}\sum_{t=1}^{N}(\hat{y}_t - y_t)^2} \ , \tag{12}$$

$$MAPE(\%) = (\frac{1}{N}\sum_{t=1}^{N}(\frac{\left|\hat{y}_t - y_t\right|}{y_t})) \cdot 100 \ , \tag{13}$$

where $y_t$ is the actual value, $\hat{y}_t$ is the predicted value, and $N$ is the number of out-of-sample data points. In this study, mean squared error (MSE) and mean absolute error (MAE) are also presented for model comparison. MSE is the squared value of the RMSE, and MAE is calculated as follows:

$$MAE = \frac{1}{N}\sum_{t=1}^{N}\left|y - \hat{y}\right|. \tag{14}$$

## 3. Data and Results

Figure 5 displays the trend of the daily electricity peak load in Korea from January 1, 2014 to October 19, 2019. Generally, the peak load has been increasing in a pattern that increases in the summer and winter, and decreases in the spring and autumn over the five-year period. In addition, the peak loads of New Year's Day and Thanksgiving are significantly less than other days; weekends and other holidays also exhibit reduced peak loads compared to weekdays.

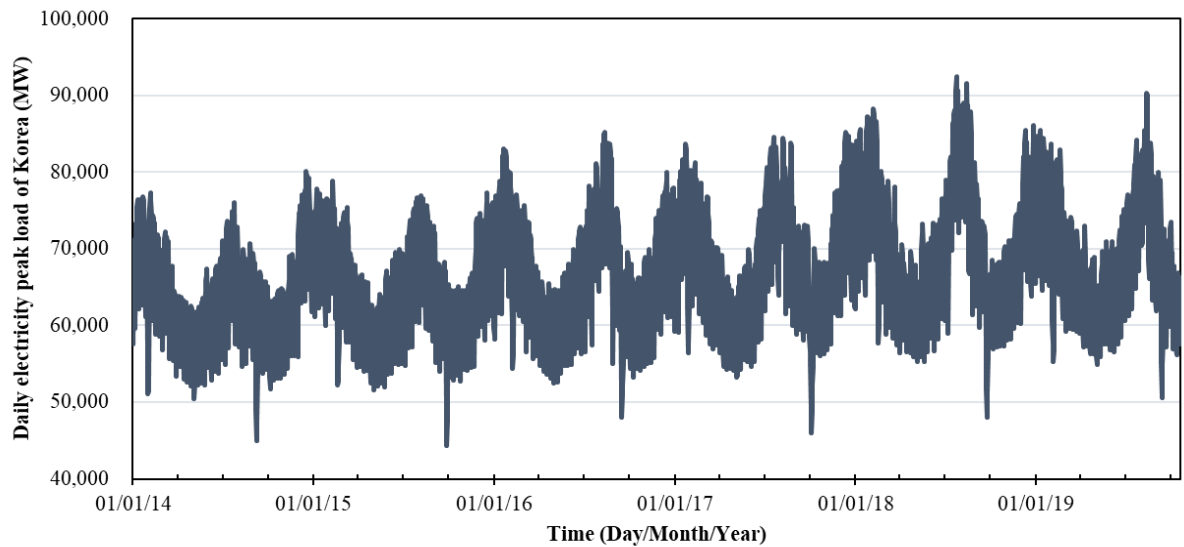

**Fig. 5.** Daily electricity peak load of Korea.

Meteorological (mean temperature and mean humidity[6]) and holiday (Saturday, Sunday, and national holidays) features are used as exogenous variables in the compared models. The histograms of the features including the peak load data are presented in Figure 6. It can be observed that the peak load has the form of a normal distribution. Moreover, because the daily mean temperature and humidity are not constant and their range is considerable, it can be confirmed that the seasonal and meteorological characteristics of Korea are distinct.

---

[6] After analyzing Pearson's correlation coefficient, this study excluded variables such as wind speed, precipitation, and cloudiness, which have relatively minimal correlation with the peak load.

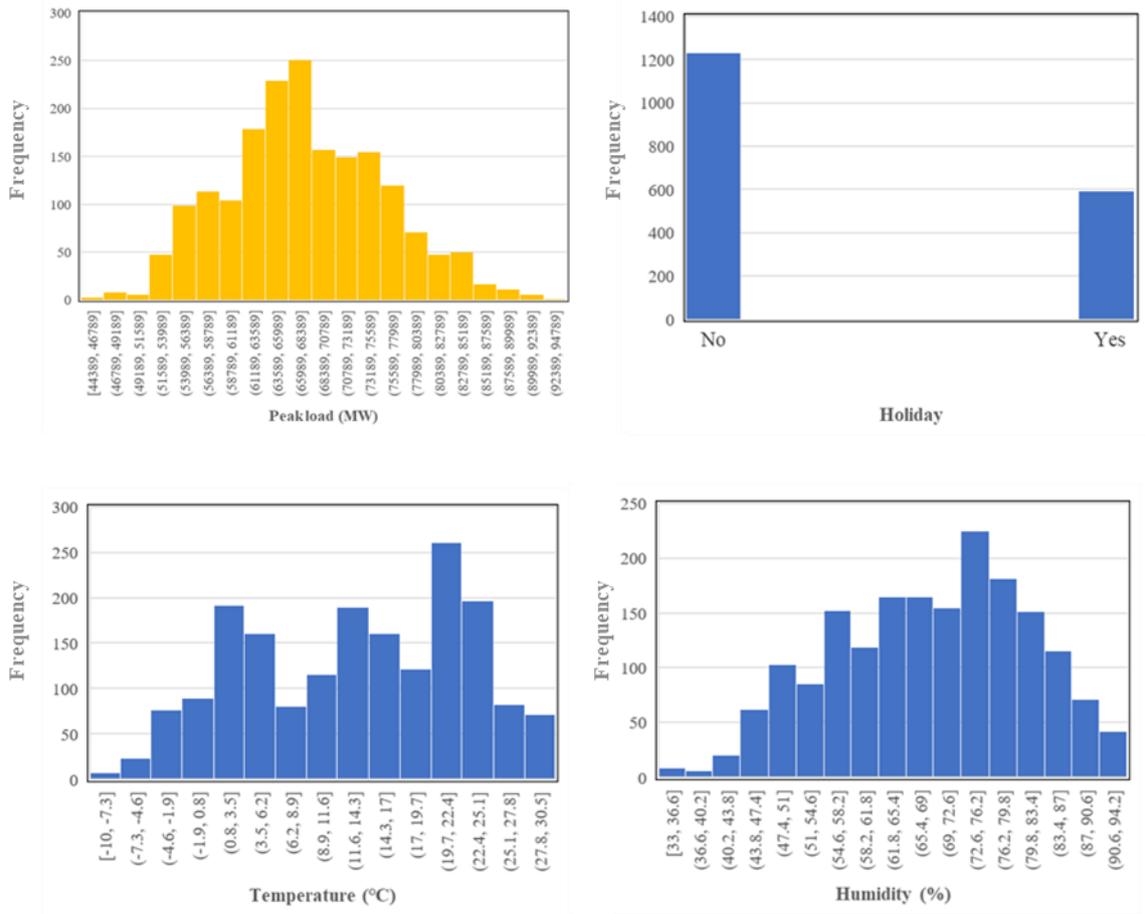

**Fig. 6.** Frequency histograms of the peak load and exogenous variables.

Of the chosen variables, the mean temperature is the most highly correlated with the peak load. However, the peak load in the summer is positively correlated with the mean temperature but negatively correlated in the winter; therefore, in this study the values of the mean temperature are squared. In addition, cooling degree day (CDD) and heating degree day (HDD), which have been considered in several previous studies as representative of the temperature effects, are also included in this study [63–66]. Given the mean temperature $T_t$, CDD and HDD can be calculated as follows:

$$CDD = \begin{cases} T_t - 24, & if \ T_t \geq 24 \\ 0, & Otherwise \end{cases}$$

$$HDD = \begin{cases} 18 - T_t, & if \ T_t \leq 18 \\ 0, & Otherwise \end{cases} \qquad (15)$$

Here, the reference temperatures[7] for the CDD and HDD of Korea were 24 ℃ and 18 ℃, respectively [67,68].

In this study, the electricity peak load data of Korea was collected from KPX [69]; the meteorological data was collected from the National Climate Data Center [70]. This study used five years of daily data from January 1, 2014 to December 31, 2018 as training data to fit the models and used test data from January 1, 2019 to October 19, 2019 to evaluate the performance (predictive power) of each model. Figure 7 displays the matrix plot of the correlation coefficient of the exogenous variables. This study included all variables because no strong (greater than 0.8) or moderate (greater than 0.6) correlations were detected [71]. Dynamic forecasts that used the previously forecasted value of the peak load when forecasting the subsequent value [72] were conducted. This study used SARIMAX (4, 1, 1) (2, 0, 0), which minimizes the AIC and BIC of the data. Moreover, hyperparameter optimization processes for the machine learning and hybrid models was conducted. This study compared the predictive powers of seven models, SARIMAX, ANN, SVR, LSTM, SARIMAX-ANN, SARIMAX-SVR, and SARIMAX-LSTM, by comparing their MSE, RMSE, MAE, and MAPE values. Because MSE and MAE are interpreted in the same manner as RMSE and MAPE, respectively, the description is based on RMSE and MAPE. When fitting the machine learning models, this study conducted a stratified five-fold cross validation process to avoid over-fitting problems.[8] K-fold cross validation is widely used to verify a model. When using this process, the dataset is repeatedly split into five folds; then the model is trained on four folds and tested on the remaining fold. This process is repeated for each fold. Step 5 in Figure 1 displays a simple graphical explanation of this process. This study utilized statistical software to manage and analyze the peak load data; SVR and ANN, including hyperparameter optimization (or grid search), were conducted using MATLAB (Mathworks) and Python; LSTM was conducted using Python; and SARIMAX and the hybrid models were conducted using R and Stata (StataCorp).

---

[7] The reference temperatures for CDD and HDD of a country are determined according to human physiological requirements, economic conditions, race, age, temperature characteristics, and other factors.

[8] For detailed processes of k-fold cross validation, please refer to Bengio and Grandvalet [73] or Rodriguez and Perez [74].

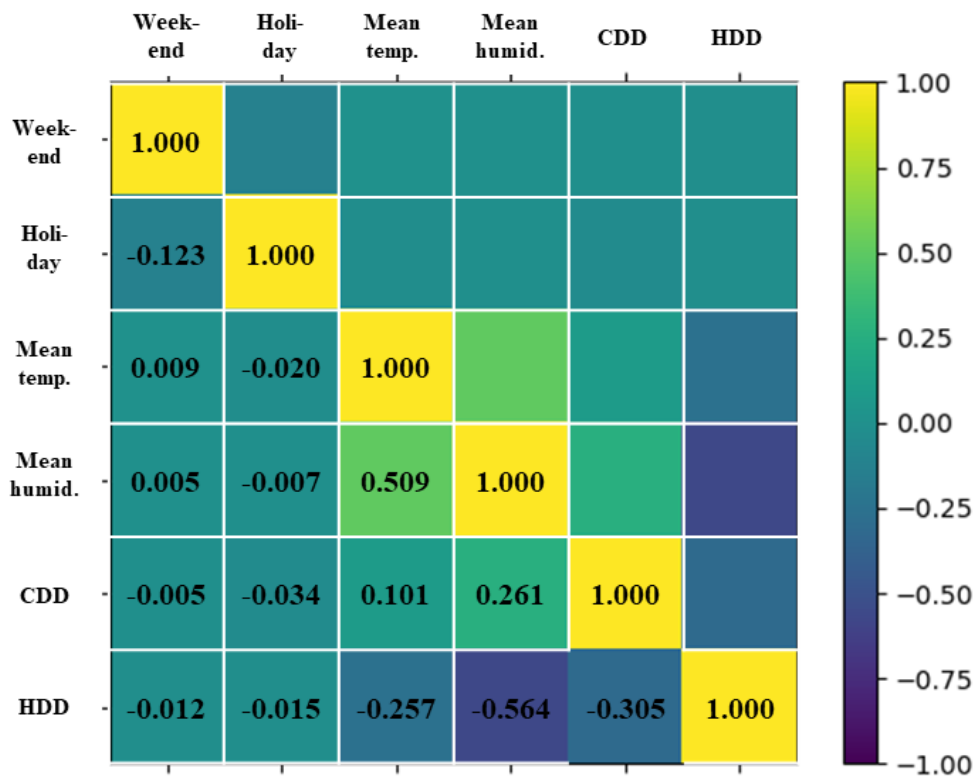

**Fig. 7.** Correlation matrix plot of exogenous variables.

## 3.1. Performance comparison

Figures 8 and 9 display the peak load prediction of the models used. In Figure 8, the orange line is the time series (SARIMAX) prediction. In Figure 9, the blue lines are the machine learning (ANN, SVR, LSTM) predictions, and green lines are the hybrid (SARIMAX-ANN, SARIMAX-SVR, and SARIMAX-LSTM) predictions. All models correctly predicted the yearly trend, indicating that the peak load is greater in the summer and winter, and less in the spring and autumn. Figure 10 presents the error distributions of the models. The forecast errors of all the models approximately followed a normal distribution (Gaussian distribution).

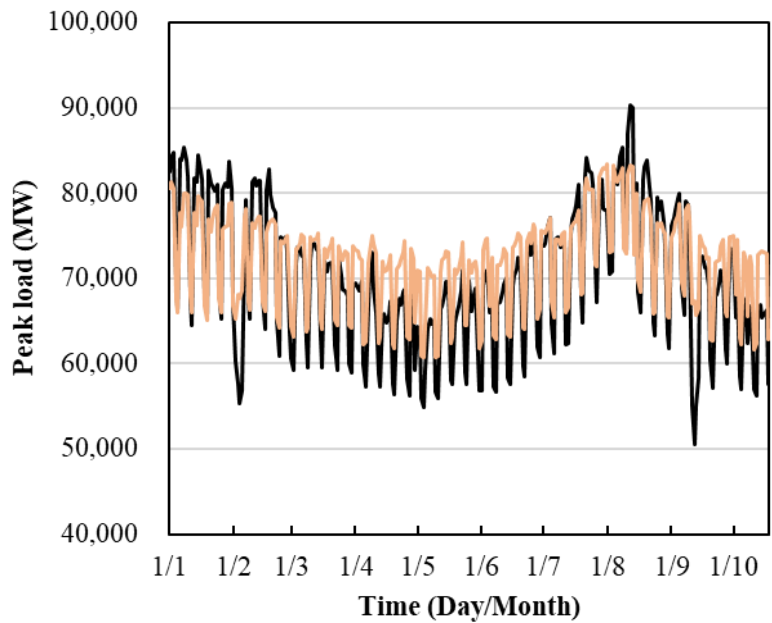

**Fig. 8.** Actual (black) and predicted (colored) peak loads of the SARIMAX model.

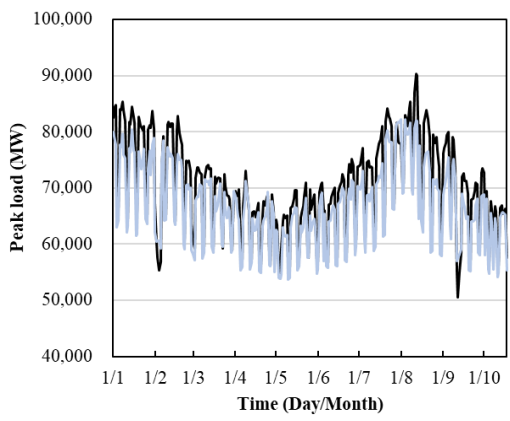

(a) ANN

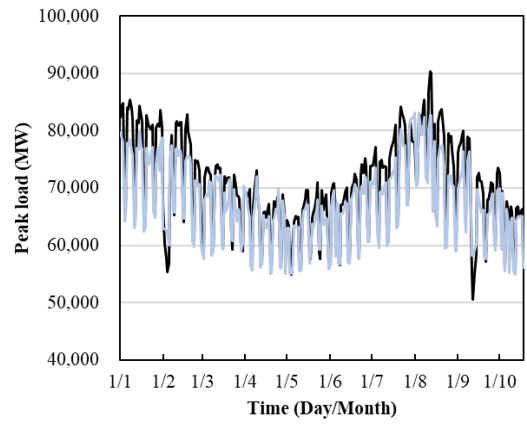

(b) SVR

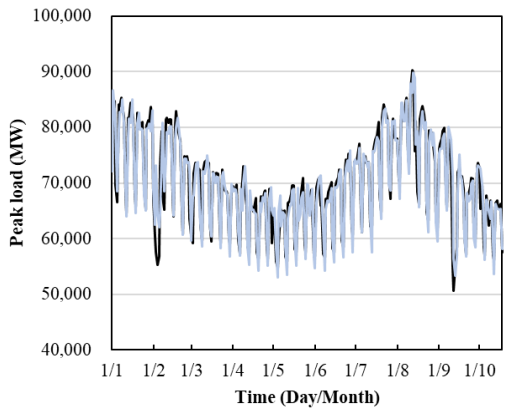

(c) LSTM

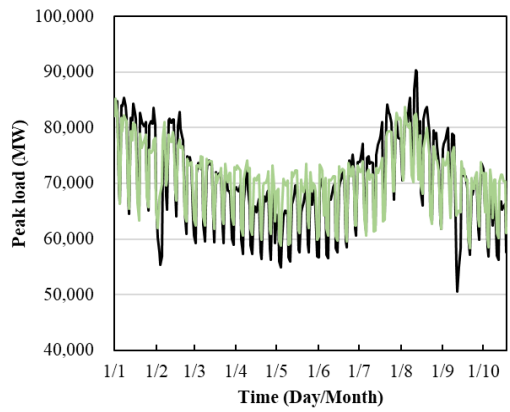

(d) SARIMAX-ANN

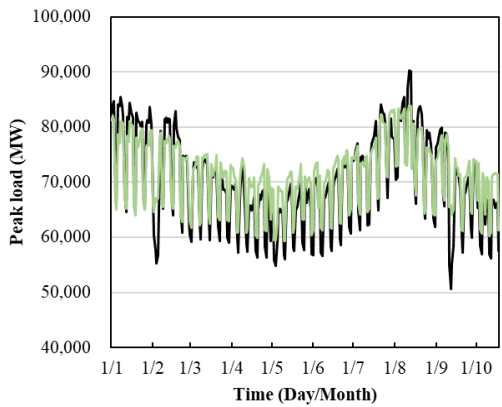

(e) SARIMAX-SVR

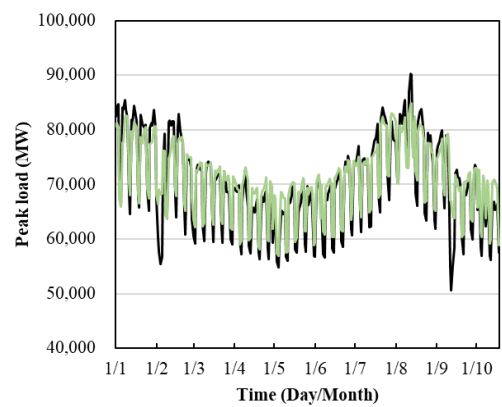

(f) SARIMAX-LSTM

**Fig. 9.** Actual (black) and predicted (colored) peak loads of machine learning and hybrid models.

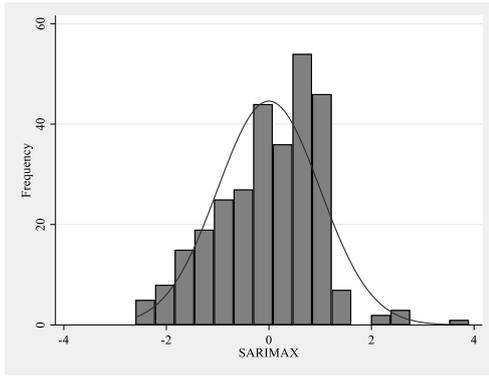

(a) SARIMAX

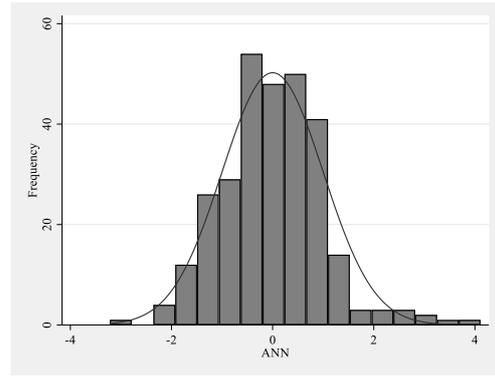

(b) ANN

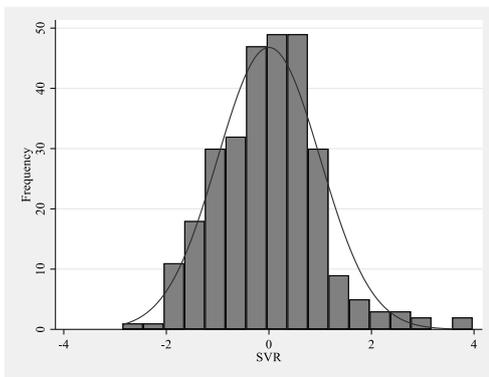

(c) SVR

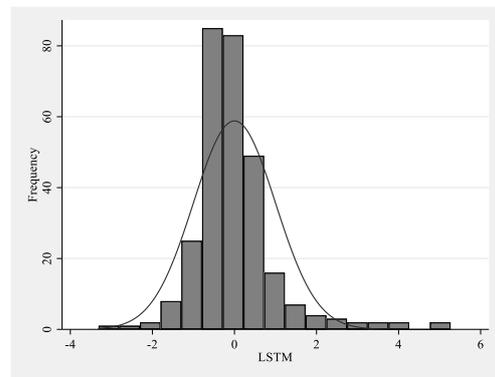

(d) LSTM

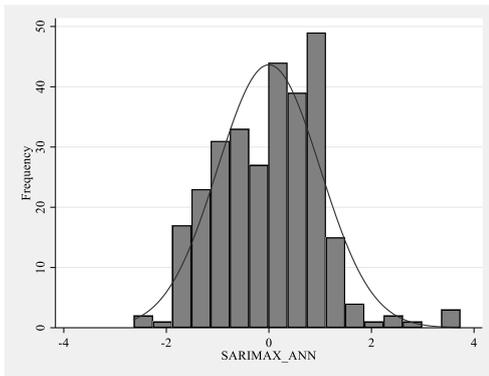

(e) SARIMAX-ANN

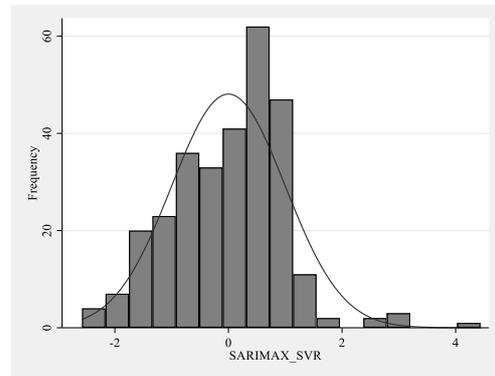

(f) SARIMAX-SVR

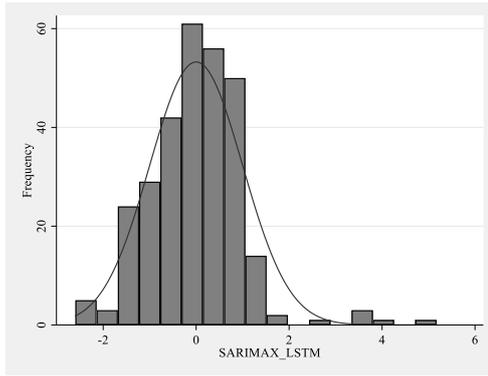

(g) SARIMAX-LSTM

**Fig. 10.** Error distribution histograms of models.

To compare each model's predictive power, this study calculated the performance evaluation factors of each model; the results are presented in Table 2.[9] When comparing single models, all machine learning models outperformed the SARIMAX model. The performance of the LSTM model was the best among the machine learning models, followed by SVR and ANN. Thus, it can be determined that the machine learning models are more suitable for peak load forecasting than the time series model in terms of predictive power.

**Table 2.** Performance evaluation factors of each model.

| Models | | MSE (GW) | MAE (MW) | RMSE (MW) | MAPE (%) | Predicted highest peak load in 2019 (MW) | Predictive $R^2$ |
|---|---|---|---|---|---|---|---|
| Single | SARIMAX | 18478.3927 | 3614.0325 | 4298.6501 | 5.4246 | 83356 | 0.796 |
| | ANN | 16889.1212 | 3562.2479 | 4109.6376 | 4.9787 | 82262 | 0.818 |
| | SVR | 13073.4383 | 3004.1915 | 3615.7210 | 4.1648 | 83116 | 0.822 |
| | LSTM | 9651.2456 | 2027.5753 | 3106.6454 | 2.9889 | 89800 | 0.861 |
| Hybrid | SARIMAX-ANN | 13036.3277 | 2943.8969 | 3610.5855 | 4.3262 | 85212 | 0.854 |
| | SARIMAX-SVR | 11552.9814 | 2788.5578 | 3398.9677 | 4.1585 | 83889 | 0.898 |
| | SARIMAX-LSTM | 9568.9936 | 2326.8125 | 3093.3790 | 3.4737 | 84811 | 0.918 |
| KSLF (KPX) | | 12468.3275 | 3088.5000 | 3531.0519 | 4.3673 | 83440 | |

*Note*: The actual highest peak of 2019 in Korea was 91300 MW.

Next, as explained in the previous section, the hybrid model was formed by combining the nonlinear parts (also called irregularity) of the time series models with the machine learning models; the training results are displayed in Figure 11. As can be observed from Table 2, all the hybrid models of this study exhibited a superior predictive power to the SARIMAX model, with the greatest improvement demonstrated by the SARIMAX-LSTM model. Moreover, when comparing the adjusted R-squared values, the goodness-of-fit of the hybrid models were all improved over the single SARIMAX model.

When comparing the LSTM and SARIMAX-LSTM models, SARIMAX-LSTM had a smaller RMSE, whereas LSTM had a smaller MAPE; however, the difference was marginal. Because this study could not identify which of the LSTM and SARIMAX-LSTM models was superior by simply comparing the accuracy-measuring criteria of this study, an additional comparison of the prediction data could be helpful. Thus, the following subsection examines the prediction power of each model in terms of the highest peak load for Korea in 2019.

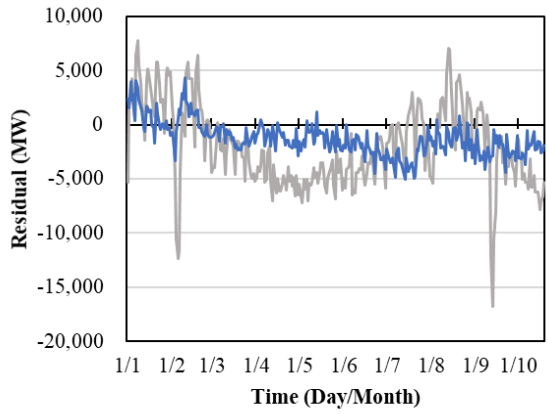

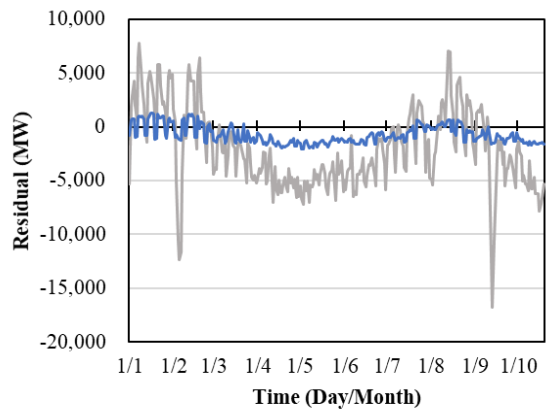

(a) ANN                                       (b) SVR

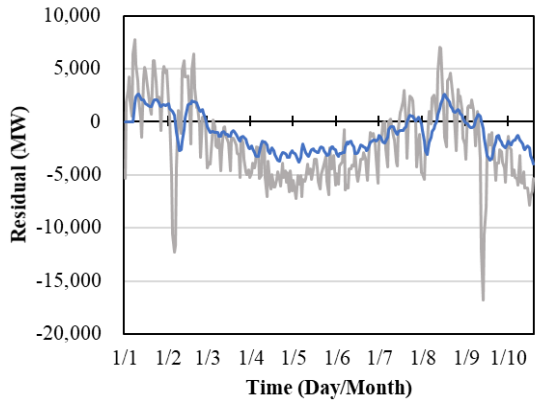

(c) LSTM

**Fig. 11.** Training results of residual of SARIMAX (gray) with machine learning models (blue).

## 3.2. The highest peak load of test data

The predictive powers of the models in this study were relatively low on days when the peak loads sharply increased or decreased. The greatest 2019 peak load was observed on August 13, at 91300 MW. This highest peak load could not be explained by the temperature, humidity, or holiday variables, which are considered in the models in this study. In Korea, the period from the end of July to early August is a nationwide peak season for summer vacation, which is not classified as a national holiday [75,76], and August 13 was the day when the majority of people started to return home and to work after the holiday season. Thus, these factors and the sweltering heat, combined to significantly boost the electricity demand, recording the highest peak load in 2019. Fortunately, KPX secured a supply capacity of approximately 96400 MW; therefore, the sharp increase in the peak load could be replenished by the supply reserve margin, avoiding supply and demand instability such as power outages (also called "blackouts"). Because damage from power outages have significant negative effects on the national economy [77–79], accurate forecasts of both general trends and abnormal increases in the daily peak loads are essential to ensure a sufficient supply capacity. In this study, the prediction of the greatest peak load in 2019 was 89800 MW, predicted by the LSTM model, which significantly outperformed all the other models used. In conclusion, the predictive power of the models covered in this study was the greatest for the SARIMAX-LSTM hybrid model and LSTM single model. The LSTM model outperformed the other models with regard to the transient changes that could not be explained by meteorological, holiday, or weekend variables.

## 3.3. Comparison with Korea's current forecasting model

This study conducted an additional analysis to compare the predictive power of the models used with the KSLF model, which KPX currently utilizes. The KSLF model consists of a panel macroeconomic model and auxiliary time series models; the model calibrates predictions for the peak load of the following day by continuously reflecting the temperature volatility of the current peak load [80,81]. A comparison between the predicted and actual values of the KSLF model is displayed in Figure 12. As shown, the KSLF model's predicted values were typically less than the actual values. Furthermore, as indicated in the last row in Table 2, the actual highest peak load in 2019 was 91300 MW; however, the KSLF model predicted a value of 83440 MW, which is less than the predictions made by LSTM and all the hybrid models of this study. The RMSE of the KSLF model was 3531.0519, and the MAPE was 4.37%; thus, the LSTM, SARIMAX-SVR, and SARIMAX-LSTM models outperformed the KSLF model in both the overall performance and greatest peak load prediction. Therefore, the performance of the current peak load-forecasting model used in Korea, which consists of panel and time series models, can be improved by including machine learning or hybrid models.

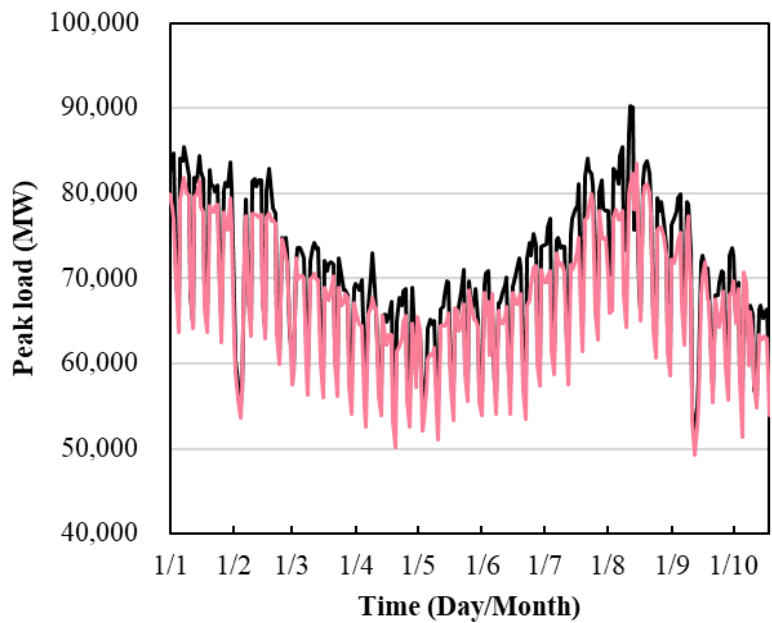

**Fig. 12.** Actual (black) and predicted (red) peak loads of the KSLF model.

## 4. Conclusions

This study comprehensively compared the performances of peak load-forecasting models, including the traditional time series model and most recent hybrid models. The seasonal autoregressive integrated moving average with exogenous variables (SARIMAX) model, artificial neural network (ANN), support vector regression (SVR), and long short-term memory (LSTM) for the single models, and SARIMAX-ANN, SARIMAX-SVR, and SARIMAX-LSTM for the hybrid models were chosen. This study successfully fits the daily electricity peak load to all models, considering weekends, holidays, and meteorological variables such as temperature, cooling and heating degree days, and humidity. The models are listed in descending order of the RMSE as follows: SARIMAX-LSTM (3093.3790), LSTM (3106.6454), SARIMAX-SVR (3398.9677), SARIMAX-ANN (3610.5855), SVR (3615.7210), ANN (4109.6376), and SARIMAX (4298.6501). The RMSE of Korea's current peak load-forecasting model was 3531.0519.

The main findings based on the forecasting results are summarized below:

(1) All single (ANN, SVR, LSTM) and hybrid (SARIMAX-ANN, SARIMAX-SVR, SARIMAX-LSTM) machine learning models exhibited significantly superior predictive power compared to the time series model (SARIMAX).

(2) LSTM and SARIMAX-LSTM exhibited the best performance among single models and hybrid models, respectively; thus, the LSTM-based models were, comparatively, the most accurate for electricity peak load forecasting in Korea.

(3) LSTM and SARIMAX-LSTM indicated no significant performance difference; however, LSTM demonstrated a superior predictive power for a sharp increase and decrease of the peak load.

(4) LSTM and SARIMAX-LSTM, which are the most predictive among the models used in this study, outperformed the current peak load-forecasting model of Korea. Therefore, Korea's current forecasting models could be significantly improved by including LSTM-based models.

Despite the achievements of this study, there are limitations. First, this study conducted electricity peak load forecasting for Korea only. As mentioned above, Korea has four distinct seasons, where peak loads increase significantly in summer and winter. Although this study observed that the single LSTM model outperformed the other models, it is not proven that the LSTM model would demonstrate the best predictive power in countries that have relatively small temperature or meteorological variations over the course of a year. Thus, it is inappropriate to apply the results of this study to all countries; an additional comparative analysis using the data from each country would be necessary.

Secondly, this study fit the learning data using tree models, Gaussian process regression and ensemble (boosted trees and bagged trees), and SVR. However, the RMSE values of these models were significantly less than that of SVR; therefore, this study did not apply them to the peak load forecasting. In the case of the hyperparameter optimization techniques, there are studies concluding that metaheuristic techniques such as the grasshopper optimization technique, whale optimization technique, ant lion optimization, and spider monkey optimization are superior to the grid search [82–84]. However, they have not been sufficiently verified through comprehensive comparative studies, and other studies have indicated that the optimization performance of grid search is superior to that of the other optimization techniques [85,86]. The optimization techniques are primarily used for classification or clustering methodologies (such as support vector machine), rather than SVR. Comprehensive and comparative studies considering different newly proposed models and techniques could be conducted in future research. However, one must preemptively analyze whether new models are applicable to peak load forecasting, have no fatal disadvantages, and can reflect the distinct seasonal and holiday characteristics of the targeted countries. Despite these limitations, this study is academically and socially meaningful because, to the best of our knowledge, it is the first attempt to successfully consider time series, machine learning, deep learning, and hybrid forecasting models for the nationwide electricity peak load.

Based on the conclusions and limitations of this study, there are meaningful future study suggestions. First, because the hybrid models combine the nonlinear residuals of a time series model by fitting them with machine learning models, the predictive power of the hybrid models is highly dependent on how well the time series model fits the linearity of the peak load data. Thus, it is important to choose a well-fitting time series model, or machine learning or other nonlinear fitting models, to develop hybrid models that can produce accurate predictions. Recently, several extended and modified SARIMAX models such as two-way SARIMAX [87] and interaction SARIMAX [88], have been proposed; they could be considered for the improvement of the predictive power of hybrid models in future studies.

Second, the highest peak load in Korea in 2019 did not exceed the supply reserve margin, and Korea could avoid a national crisis, such as a blackout. However, as abnormal weather caused by climate change is becoming more frequent, Korea's current forecasting models must be improved. Moreover, further studies to derive an optimal supply reserve margin for the supply and demand balance could be conducted based on accurate peak load forecasting.

Third, this study selected the forecasting models that were the most widely used in previous studies; hence, it did not consider the latest extended models. As mentioned earlier, LSTM is one of the extended models used to overcome the shortcomings of RNN models. However, ANN- [89–91] and SVM-extended models [92–94] have also been proposed and utilized, with a variety of derivative models complementing the disadvantages of basic models. In addition to the neural network or support vector-based algorithms chosen in this study, the boosting algorithm (typically XGboost) is also being used actively, and its high performance has been validated. Thus, it is also necessary to compare the forecasting performance by adopting the boosting and boosting-hybrid models for future peak load forecasting.

Lastly, this study conducted peak load forecasting on a daily basis and compared the predictive powers of the different models. However, for electricity demand forecasting, electricity consumption[10] or peak load occurrence time could also be analyzed with the peak load and a varying analysis period from the short term (hours) to long term (weeks, months, and seasons). Therefore, in the future, to compare the performance of the models used in this study, an electricity consumption forecasting analysis based on different time units will be necessary. For example, when conducting the peak load-forecasting studies targeting working days, utilization rates, or volumes of machinery, appliances and bulbs could be reflected as exogenous variables influencing the peak load. In addition, when conducting studies targeting specific seasons such as midwinter or midsummer, it could be helpful to consider the maximum and minimum temperature rather than the mean temperature only to reflect the climate fluctuation.

---

[10] The unit of electricity load is W, and the unit of electricity consumption is Wh.

**Funding**

This research did not receive any specific grant from funding agencies in the public, commercial, or not-for-profit sectors.